\documentclass[%
 aip,
 jmp,%
 amsmath,amssymb,
reprint,%
]{revtex4-1}

\usepackage{graphicx}
\usepackage{dcolumn}
\usepackage{bm}
\usepackage{hyperref}

\begin{document}

\title{Fabry-P\'erot interference in a triple-gated quantum point contact}

\author{S. Maeda}
 \email{s.maeda.phys@gmail.com}
\author{S. Miyamoto}
 \altaffiliation[Present address: ]{School of Fundamental Science and Technology, Keio University, 3-14-1 Hiyoshi, Kohoku-ku, Yokohama 223-8522, Japan
 }
\author{M. H. Fauzi}
\author{K. Nagase}
\author{K. Sato}
\author{Y. Hirayama}
 \email{hirayama@m.tohoku.ac.jp}
\affiliation{
 Department of Physics, Tohoku University, 6-3 Aramaki Aza Aoba, Aoba-ku, Sendai 980-8578, Japan
}

\date{\today}

\begin{abstract}
We demonstrated that a triple-gated GaAs quantum point contact, which has an additional surface gate between a pair of split gates to strengthen the lateral confinement, produces the well-defined quantized conductance and Fabry-P\'erot-type (FP-type) oscillations on it even using a relatively low mobility wafer. A one-dimensional phenomenological model potential was developed to explain the oscillatory behavior. By combining the model calculations and dc bias spectroscopy, we obtained detailed information about the energy scales of the oscillatory structures. The relationships between the FP-type oscillations and the anomaly below the first plateau will be addressed.
\end{abstract}

\maketitle

Electronic analogs of optical interferometric devices realized in semiconductor mesoscopic systems are attracting attention owing to their controllability and scalability\cite{ji:2003nature,yamamoto:2012naturenano}. For such solid-state quantum devices, a quantum point contact (QPC), a short and narrow constriction between two electron reservoirs, is frequently used as a waveguide that selectively transmits the propagating electrons. The QPC is formed in a two-dimensional electron system (2DES) by depositing a pair of surface split gates opposite each other, and consequently its conductance $G$ is quantized in units of $G_0 = 2e^2/h$ as a function of the split gate bias\cite{wees:1988prl,wharam:1988jphysc}.

The QPC itself can act as a resonator based on quantum interference. One of the possible applications of this QPC resonator is a switching device where resonance and antiresonance features on the quantized conductance can be identified as on and off of the switch\cite{flory:1997jap}. Conductance oscillations due to the partial reflections at both ends of the constriction were theoretically predicted in the early studies of the QPC transport\cite{szafer:1989prl,marel:1989prb,tekman:1991prb}. However, such oscillations have hardly been observed due to the adiabatic connection between the one-dimensional (1D) channel and the 2DES reservoirs\cite{yacoby:1990prb}. To obtain the conductance oscillations, several QPC structures have been proposed and tested. For instance, electron stub tuners (T-shaped waveguide)\cite{sols:1989apl,debray:2000prb}, successive narrow and wide structures\cite{kouwenhoven:1990prl,frost:1994prb}, structures integrated with a cross gate\cite{wen:2010prb}, or waveguides defined with the etching\cite{wang:2000apl} presented observable conductance oscillations. Unlike the conventional semiconductor devices that operate in a diffusive regime, these QPC resonators based on quantum interference easily suffer from the weak disorders such as small defects or impurities in the device. Therefore, it is important to clarify what kind of factor is important to enhance the visibility of Fabry-P\'erot-type (FP-type) interference oscillations. In this context, it is noteworthy that the FP-type oscillation has been observed in carbon nanotubes\cite{liang:2001nature} and InAs nanowire\cite{kretinin:2010nanolett} in spite of the lower mobility than that of the GaAs high mobility systems. The unique points of these systems are prominent scattering at the nanowire-electrode interface and a relatively uniform wire region in the center.

In this letter, we report that a simple triple-gated QPC device produces the well-defined FP-type oscillations on the conductance plateaus even using a relatively low mobility wafer. This arises from uniform wire potential at the center and enhanced reflection at the edge created by the triple-gate structure with an appropriate bias condition. A minimal phenomenological model was developed to explain the behavior of the oscillations and used to characterize the device properties. The clear quantization and the FP-type oscillation in GaAs triple-gated QPC enables us to access interplay between the $< G_0$ anomaly and the FP-type oscillations.

\begin{figure}
    \centering
    \includegraphics[scale=1]{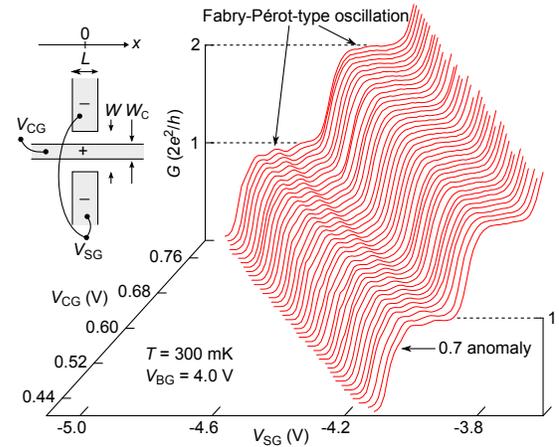}
    \caption{Main panel plots linear conductance $G$ at $T = 300$ mK as function of $V_{\mathrm{SG}}$ for different $V_{\mathrm{CG}}$. $V_{\mathrm{CG}}$ was increased incrementally from 0.4 to 0.8 V in steps of 0.01 V (traces are not offset). Left-side panel schematically illustrates triple-gate layout used in experiment. $x$ axis is chosen to be parallel to transport direction. Nominal gate dimensions are $L = 400$ nm, $W = 500$ nm, and $W_\mathrm{C} = 200$ nm.}
    \label{fig1}
\end{figure}
We used two relatively low-mobility devices with the triple-gate layout\cite{chou:1993apl,lee:2006jap} shown in the left-side of Fig.~\ref{fig1} to define a 1D transport channel. The nominal gate dimensions are $L = 400$ nm, $W = 500$ nm, and $W_\mathrm{C} = 200$ nm. The 2DES located 140 nm below a GaAs/AlGaAs heterostructure surface shows the mobility of 30 $\mathrm{m^2/Vs}$ for an electron density $n_\mathrm{2D}$ of $2.0\times10^{15}\ \mathrm{m}^{-2}$ at $T =$ 1.7 K. A dominant scattering mechanism in the 2DES can be estimated by comparing a quantum lifetime $\tau_\mathrm{q}$ calculated from the Shubnikov$\mathchar`-$de Haas oscillations and a transport lifetime $\tau_\mathrm{t}$ obtained from the mobility\cite{harrang:1985prb}. The ratio of both the lifetimes $\tau_\mathrm{t}/\tau_\mathrm{q}$ at $n_\mathrm{2D} = 2.0\times10^{15}\ \mathrm{m}^{-2}$ was approximately 11 for our modulation-doped back-gated\cite{hamilton:1992apl} device. According to MacLeod $\it{et\ al.}$\cite{macleod:2009prb}, $\tau_\mathrm{t}/\tau_\mathrm{q}\ \protect\raisebox{-0.5ex}{$\:\stackrel{\textstyle <}{\sim}\:$} (\protect\raisebox{-0.5ex}{$\:\stackrel{\textstyle >}{\sim}\:$})\ 10$ implies that the background (remote) impurities predominantly contribute to the scattering events, so in our case, both the mechanisms can play a significant role in the scatterings. A measurement was carried out in a cryofree $^3\mathrm{He}$ refrigerator, and a differential conductance $G = dI/dV$ at $T = 300$ mK was recorded using a standard four-terminal lock-in technique. The series resistance from the bulk 2DES was subtracted for the presented data.

The main panel of Fig.~\ref{fig1} plots conductance traces of a typical low-mobility device as a function of split-gate voltage $V_\mathrm{SG}$ for different center-gate bias $V_\mathrm{CG}$. A positive bias applied to the center gate increases the 1D subband spacing and the electron density underneath. Consequently, the mode mixing is suppressed and the fluctuated potential is screened, leading to the well-defined quantization even for our relatively low-mobility devices. The anomalous $0.7G_0$ shoulder-like plateau\cite{thomas:1996prl}, which is usually attributed to the many-body effect, appeared below the $G_0$ plateau. We also observed the FP-type conductance oscillations on the conductance plateaus. The FP-type oscillations become stronger as $V_\mathrm{CG}$ increases, which indicates that the origin is not disorder but the gate-geometrical effect. We consider that the saddle-point potential induced by the split gate is modified by the center gate, and the barrier top becomes flatter than parabolic\cite{heyder:2015prb}, consequently generating the FP-type oscillations. Note that such conductance oscillations were also observed in our high-mobility triple-gated device (100-200\ $\mathrm{m^2/Vs}$). Similar oscillations were reported by Reilly $\it{et\ al.}$\cite{reilly:2001prb}, too; however, longer $L$ of 2 $\mu$m makes the origin of the oscillatory features ambiguous because of the not negligible effect of a background random potential\cite{smith:2016prappl}. In addition, interestingly enough, the 0.7 anomalous structure clearly remains as it is after the development of the FP-type oscillations in Fig.~\ref{fig1}.

We developed a simple phenomenological 1D model to explain the experimental observations. 
\begin{figure}
    \centering
    \includegraphics[scale=1]{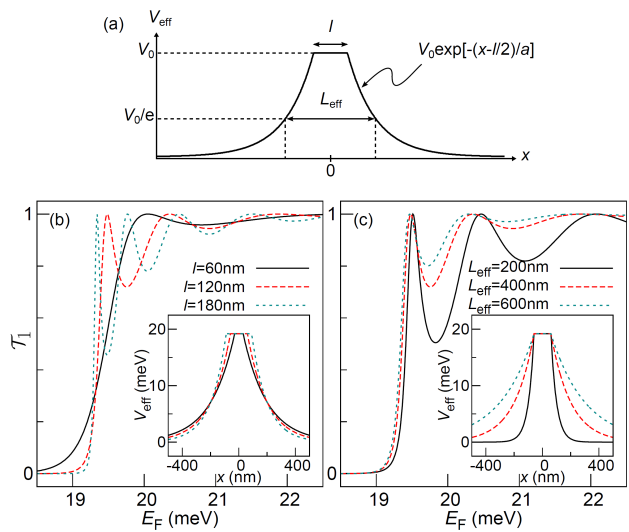}
    \caption{(a) Modeled 1D effective potential barrier that has flat top region of length $l$ and exponentially decaying slope on either side of flat region ($a$: decay constant of potential, $L_\mathrm{eff} = l+2a$: effective channel length). (b), (c) Calculated $E_\mathrm{F}$-$\mathcal{T}_1$ traces for fixed (b) $L_\mathrm{eff}$ (= 400 nm) and (c) $l$ (= 120 nm); corresponding $V_\mathrm{eff}$ is shown in insets.}
    \label{fig2}
\end{figure}
Figure \ref{fig2}(a) shows a modeled 1D effective potential barrier, which has a flat top region of length $l$ and an exponentially decaying slope on either side of the flat region:
\begin{eqnarray}
V_\mathrm{eff}(x)=\left\{ \begin{array}{ll}
V_0 \hspace{92pt} |x|\leq l/2\\
V_0\mathrm{exp}[-(x-l/2)/a] \hspace{20pt} x>l/2\ \ \ ,\\
V_0\mathrm{exp}[(x+l/2)/a] \hspace{28pt} x<-l/2\\
\end{array} \right.
\end{eqnarray}
where $V_0$ is the barrier height and $a$ is a decay constant of the potential. The potential is reduced to $V_0$/e (e: Euler's number) at $|x| = l/2+a$; we introduce an effective channel length $L_\mathrm{eff} = l+2a$ as a characteristic length. The independent parameters of this model are therefore $V_0$, $l$, and $L_\mathrm{eff}$.

The transmission for the lowest 1D subband $\mathcal{T}_1$ was calculated by solving 1D Schr\"odinger equation for fixed barrier height $V_0 = 19.2$ meV. This height is an estimated Fermi energy $E_\mathrm{F}$ under the center gate far from the QPC center for $V_\mathrm{BG} = 4.0$ V and $V_\mathrm{CG} = 0.8$ V. The transmission for the $n$th 1D subband $\mathcal{T}_n$ rerates to the conductance $G$ at finite dc bias $V_\mathrm{sd}$ as
\begin{equation}
G = -\frac{e^2}{h}\sum_n \int dE\mathcal{T}_n(E)[f^\prime(E+\frac{eV_\mathrm{sd}}{2})+f^\prime(E-\frac{eV_\mathrm{sd}}{2})],
\end{equation}
where $f^\prime(E)$ is a derivative of the Fermi distribution $f(E) = [1+\mathrm{exp}(-E/k_\mathrm{B}T)]$ with temperature $T$. Thus the linear conductance at zero temperature is given by $G=G_0\sum_n\mathcal{T}_n(E)$. Figures \ref{fig2}(b) and (c) show the calculated $E_\mathrm{F}$-$\mathcal{T}_1$ traces for the fixed $L_\mathrm{eff}$ (= 400 nm) and $l$ (= 120 nm), respectively; the corresponding $V_\mathrm{eff}$ is shown in the insets. Note that we fix the chemical potential and vary the barrier size and shape in practice. The energy intervals between the resonance peaks are less affected by the $L_\mathrm{eff}$ while they strongly depend on $l$, indicating the importance of the flat top region. We can extract the $l$ by comparing the calculation with the experiment. All the results including a dc bias experiment mentioned later have been reproduced for $l = 120$ nm. In contrast, it is difficult to precisely estimate the $L_\mathrm{eff}$. The $L_\mathrm{eff}$ primarily affects the strength of the resonance peaks; the small $L_\mathrm{eff}$ makes the peak pattern strong and vice versa. In particular, the higher resonance peaks are more affected by the $L_\mathrm{eff}$, and as shown in Fig.~\ref{fig2}(c), the second resonance peak with the short Fermi wavelength rapidly smears with increasing the $L_\mathrm{eff}$ from 200 to 400 nm. In the experiment, however, the conductance curve shows the observable second resonant peak. Considering the above, we can roughly estimate a maximum of the $L_\mathrm{eff}$; $L_\mathrm{eff} \protect\raisebox{-0.5ex}{$\:\stackrel{\textstyle <}{\sim}\:$} 400$ nm. Although the sharp corners make the oscillatory patterns strong, potential discontinuities as assumed in our model are not needed to produce the oscillations. 
\begin{figure}
    \centering
    \includegraphics[scale=1]{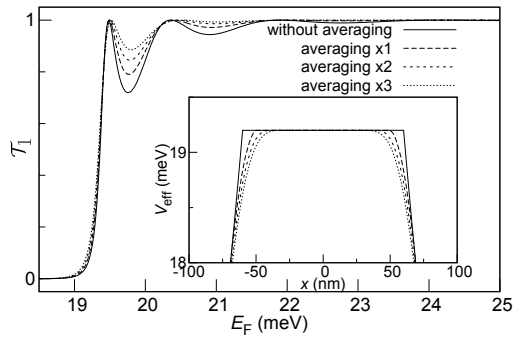}
    \caption{$E_\mathrm{F}$-$\mathcal{T}_1$ traces where each data point of potential is replaced with average of five adjacent data points. Note that distance between two successive data points is set to 2 nm in present calculations. Each trace is result of zero-three times averaging; corresponding $V_\mathrm{eff}$ is shown in inset. Here $l$ and $L_\mathrm{eff}$ are fixed to 120 and 400 nm, respectively.}
    \label{fig3}
\end{figure}
Figure \ref{fig3} shows $E_\mathrm{F}$-$\mathcal{T}_1$ traces when the sharp corner edges in the model potential are rounded. Though the FP-type oscillations are weakened by the rounding, they clearly remains without the sharp corner.
We simply assume $L_\mathrm{eff} = L = 400$ nm and no rounding in the following discussions.

A dc bias measurement provides detailed information about the energy scales of 1D transport phenomena. 
\begin{figure*}
    \centering
    \includegraphics[scale=1]{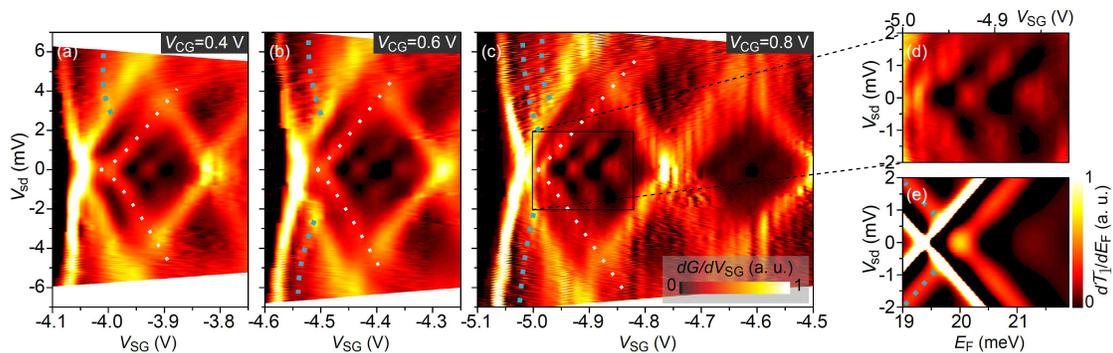}
    \caption{(a)-(c) Color plots of $dG/dV_\mathrm{SG}$ as functions of $V_\mathrm{SG}$ and $V_\mathrm{sd}$ for $V_\mathrm{CG}$ = 0.4, 0.6, and 0.8 V, where peaks in $dG/dV_\mathrm{SG}$ appear as bright lines. White dotted line highlights anomalous 1D subband edge between 0.8-0.85$G_0$ finite bias plateau and $G_0$ plateau. (d) Enlarged view of checkerboard pattern in (c). (e) Calculated $d\mathcal{T}_1/dE_\mathrm{F}$ assuming $l$ = 120 nm and $V_0$ = 19.2 meV. Blue dotted lines highlight peak lines below $G_0$ plateau. These lines also appeared in (a)-(c) (highlighted by blue dotted lines).}
    \label{fig4}
\end{figure*}
Figures \ref{fig4}(a)-(c) show color plots of the transconductance $dG/dV_\mathrm{SG}$ as functions of $V_\mathrm{SG}$ and source-drain bias $V_\mathrm{sd}$ for $V_\mathrm{CG}$ = 0.4, 0.6, and 0.8 V, respectively, where the peaks in $dG/dV_\mathrm{SG}$ appear as bright lines. The large bright diamonds correspond to the conditions on which the 1D subband edges cross the chemical potential. The white dotted lines highlight the anomalous 1D subband edge between the 0.8-0.85$G_0$ finite bias plateau and the $G_0$ plateau. It is widely known that the 0.7 anomaly evolves into a 0.8-0.85 plateau under the finite bias\cite{kristensen:2000prb}. On the $G_0$ plateau, small checkerboard-like patterns appeared to be associated with the FP-type resonances seen in Fig.~\ref{fig1}. The vertical size of the checkerboard ($\it{i.e.}$, energy intervals between the successive resonance peaks) barely depends on the $V_\mathrm{CG}$, which suggests the flat top length $l$ is almost constant in the $V_\mathrm{CG}$ range.

We compared the observed checkerboard with a calculation using our model described in Fig.~\ref{fig2}(a). Figure \ref{fig4}(d) shows an enlarged view of the checkerboard pattern in Fig.~\ref{fig4}(c). As shown in Fig.~\ref{fig4}(e), a calculated $d\mathcal{T}_1/dE_\mathrm{F}$ assuming $l = 120$ nm and $V_0 = 19.2$ meV quantitatively well reproduced the observed pattern. The calculation also predicted the peak lines below the $G_0$ plateau (highlighted by blue dotted lines) that they are just the extensions of the FP-type resonance peaks on the $G_0$ plateau. These lines also appeared in the experimental data (highlighted by blue dotted lines). It is noteworthy that they disappeared on the 0.8-0.85$G_0$ finite bias plateau, suggesting disappearance of the FP-type oscillation in this regime. This disappearance suggests that the spin degeneracy is lifted and/or the potential shape is modified by the enhanced electron-electron interactions.

Finally, we addressed the strength of the FP-type oscillations for the different $V_\mathrm{CG}$ and propagating modes.
Positive bias applied to the center gate lowers the potential outside the 1D channel as well as the 1D channel potential, and we practically fix the chemical potential during the measurement. Consequently, more negative $V_\mathrm{SG}$ is required for large $V_\mathrm{CG}$ conditions in order to close the 1D conducting channel. This trend is just depicted in Fig.~\ref{fig1}. Thus increasing $V_\mathrm{CG}$ means increasing $V_0$ in our model.
\begin{figure}
    \centering
    \includegraphics[scale=1]{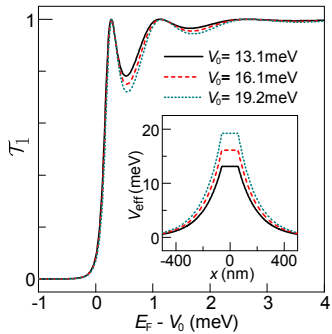}
    \caption{$(E_\mathrm{F}-V_0)$-$\mathcal{T}_1$ traces for three different $V_0$; each corresponds to estimated $E_\mathrm{F}$ under center gate far from QPC center for $V_\mathrm{CG}$ = 0.4, 0.6, and 0.8 V. Here $l$ and $L_\mathrm{eff}$ are fixed to 120 and 400 nm, respectively.}
    \label{fig5}
\end{figure}
Figure \ref{fig5} shows the calculated $(E_\mathrm{F}-V_0)$-$\mathcal{T}_1$ traces for the three different $V_0$; each corresponds to an estimated $E_\mathrm{F}$ under the center gate far from the QPC center for $V_\mathrm{CG}$ = 0.4, 0.6, and 0.8 V. Here $l$ and $L_\mathrm{eff}$ are fixed to 120 and 400 nm, respectively. The result shows that the FP-type oscillations become stronger as $V_0$ ($\it{i.e.}$, $V_\mathrm{CG}$) increases. Hence the observed dependence of the resonance patterns on the $V_\mathrm{CG}$ is partly attributed to the change of the $V_0$. As shown in Figs.~\ref{fig1} and \ref{fig4}(c), the FP-type oscillations on the $2G_0$ plateau are much weaker than those on the $G_0$ plateau even though the $V_0$ is the same for different propagating modes. We infer that the two corner edges of the flat top region were rounded by the screening effect due to the electrons propagating in the first mode, which consequently weakens the FP-type resonances on the second plateau. Obviously, the change of $L_\mathrm{eff}$ for the different modes can also contribute to the resonance strength.

We summarize here the criteria to obtain clear FP-type oscillations on the conductance plateau. As shown in Figs.~\ref{fig2}(b), \ref{fig2}(c), \ref{fig3}, and \ref{fig5}, high and steep potential barrier with the sharp corner edges ($\it{i.e.}$, abrupt transition from the reservoir to the channel) is required to enhance visibilities of the oscillations. Hence etched waveguides or nanowires are more suitable than surface-gated QPCs in order to observe the clear FP-type oscillations while they have less controllability compared with the surface-gated QPCs. From the measuring side, of course the lower temperature and less noise are preferred. Considering the above, length $l$ of the flat top should be appropriate dimension for the specific device and measuring condition used. Small $l$ results in weak oscillations with the large interval, and oscillations with the small interval which correspond to large $l$ is easily suffer from the noise and energy averaging. The channel length dependence of the FP-type oscillations have been studied using the etched waveguides\cite{wang:2000apl} or the nanowires\cite{liang:2001nature} with the different channel length. By contrast, in the surface-gated systems, there is a possibility to realize a length tunable QPC resonator, for example by combining the present triple-gated structure with a length tunable QPC reported by Iqbal $\it{et\ al.}$\cite{iqbal:2013jap}.

In conclusion, we demonstrated that a simple triple-gated QPC device produces well-defined quantized conductance and the FP-type oscillations on it even using a relatively low-mobility wafer. A dc bias measurement provided detailed information about the energy scales of the oscillatory structures. A minimal phenomenological model was developed to explain the oscillatory behavior. By combining the model calculations and dc bias spectroscopy, we characterized the device properties, such as a resonant length. Finally, our experiments indicated that the FP-type oscillations did not disturb the 0.7 anomaly at zero bias and, more interestingly, disappeared in the 0.8-0.85 plateau region under the finite bias condition.

We greatly appreciate the support from ERATO Nuclear Spin Electronics Project, JST. S.M., M.H.F. and Y.H. acknowledge support from Interdepartmental Doctoral Degree Program for Multi-dimensional Material Science Leaders at Tohoku University. K.N. and Y.H. acknowledge support from Graduate Program in Spintronics, Tohoku University. Y.H. acknowledge financial support from KAKENHI Grant No.~26287059 and No.~15H05867.


\begin{thebibliography}{28}%
\makeatletter
\providecommand \@ifxundefined [1]{%
 \@ifx{#1\undefined}
}%
\providecommand \@ifnum [1]{%
 \ifnum #1\expandafter \@firstoftwo
 \else \expandafter \@secondoftwo
 \fi
}%
\providecommand \@ifx [1]{%
 \ifx #1\expandafter \@firstoftwo
 \else \expandafter \@secondoftwo
 \fi
}%
\providecommand \natexlab [1]{#1}%
\providecommand \enquote  [1]{``#1''}%
\providecommand \bibnamefont  [1]{#1}%
\providecommand \bibfnamefont [1]{#1}%
\providecommand \citenamefont [1]{#1}%
\providecommand \href@noop [0]{\@secondoftwo}%
\providecommand \href [0]{\begingroup \@sanitize@url \@href}%
\providecommand \@href[1]{\@@startlink{#1}\@@href}%
\providecommand \@@href[1]{\endgroup#1\@@endlink}%
\providecommand \@sanitize@url [0]{\catcode `\\12\catcode `\$12\catcode
  `\&12\catcode `\#12\catcode `\^12\catcode `\_12\catcode `\%12\relax}%
\providecommand \@@startlink[1]{}%
\providecommand \@@endlink[0]{}%
\providecommand \url  [0]{\begingroup\@sanitize@url \@url }%
\providecommand \@url [1]{\endgroup\@href {#1}{\urlprefix }}%
\providecommand \urlprefix  [0]{URL }%
\providecommand \Eprint [0]{\href }%
\providecommand \doibase [0]{http://dx.doi.org/}%
\providecommand \selectlanguage [0]{\@gobble}%
\providecommand \bibinfo  [0]{\@secondoftwo}%
\providecommand \bibfield  [0]{\@secondoftwo}%
\providecommand \translation [1]{[#1]}%
\providecommand \BibitemOpen [0]{}%
\providecommand \bibitemStop [0]{}%
\providecommand \bibitemNoStop [0]{.\EOS\space}%
\providecommand \EOS [0]{\spacefactor3000\relax}%
\providecommand \BibitemShut  [1]{\csname bibitem#1\endcsname}%
\let\auto@bib@innerbib\@empty
\bibitem [{\citenamefont {Ji}\ \emph {et~al.}(2003)\citenamefont {Ji},
  \citenamefont {Chung}, \citenamefont {Sprinzak}, \citenamefont {Heiblum},
  \citenamefont {Mahalu},\ and\ \citenamefont {Shtrikman}}]{ji:2003nature}%
  \BibitemOpen
  \bibfield  {author} {\bibinfo {author} {\bibfnamefont {Y.}~\bibnamefont
  {Ji}}, \bibinfo {author} {\bibfnamefont {Y.}~\bibnamefont {Chung}}, \bibinfo
  {author} {\bibfnamefont {D.}~\bibnamefont {Sprinzak}}, \bibinfo {author}
  {\bibfnamefont {M.}~\bibnamefont {Heiblum}}, \bibinfo {author} {\bibfnamefont
  {D.}~\bibnamefont {Mahalu}}, \ and\ \bibinfo {author} {\bibfnamefont
  {H.}~\bibnamefont {Shtrikman}},\ }\href@noop {} {\bibfield  {journal}
  {\bibinfo  {journal} {Nature}\ }\textbf {\bibinfo {volume} {422}},\ \bibinfo
  {pages} {415} (\bibinfo {year} {2003})}\BibitemShut {NoStop}%
\bibitem [{\citenamefont {Yamamoto}\ \emph {et~al.}(2012)\citenamefont
  {Yamamoto}, \citenamefont {Takada}, \citenamefont {B{\"a}uerle},
  \citenamefont {Watanabe}, \citenamefont {Wieck},\ and\ \citenamefont
  {Tarucha}}]{yamamoto:2012naturenano}%
  \BibitemOpen
  \bibfield  {author} {\bibinfo {author} {\bibfnamefont {M.}~\bibnamefont
  {Yamamoto}}, \bibinfo {author} {\bibfnamefont {S.}~\bibnamefont {Takada}},
  \bibinfo {author} {\bibfnamefont {C.}~\bibnamefont {B{\"a}uerle}}, \bibinfo
  {author} {\bibfnamefont {K.}~\bibnamefont {Watanabe}}, \bibinfo {author}
  {\bibfnamefont {A.~D.}\ \bibnamefont {Wieck}}, \ and\ \bibinfo {author}
  {\bibfnamefont {S.}~\bibnamefont {Tarucha}},\ }\href@noop {} {\bibfield
  {journal} {\bibinfo  {journal} {Nat. nanotechnol.}\ }\textbf {\bibinfo
  {volume} {7}},\ \bibinfo {pages} {247} (\bibinfo {year} {2012})}\BibitemShut
  {NoStop}%
\bibitem [{\citenamefont {van Wees}\ \emph {et~al.}(1988)\citenamefont {van
  Wees}, \citenamefont {van Houten}, \citenamefont {Beenakker}, \citenamefont
  {Williamson}, \citenamefont {Kouwenhoven}, \citenamefont {van~der Marel},\
  and\ \citenamefont {Foxon}}]{wees:1988prl}%
  \BibitemOpen
  \bibfield  {author} {\bibinfo {author} {\bibfnamefont {B.~J.}\ \bibnamefont
  {van Wees}}, \bibinfo {author} {\bibfnamefont {H.}~\bibnamefont {van
  Houten}}, \bibinfo {author} {\bibfnamefont {C.~W.~J.}\ \bibnamefont
  {Beenakker}}, \bibinfo {author} {\bibfnamefont {J.~G.}\ \bibnamefont
  {Williamson}}, \bibinfo {author} {\bibfnamefont {L.~P.}\ \bibnamefont
  {Kouwenhoven}}, \bibinfo {author} {\bibfnamefont {D.}~\bibnamefont {van~der
  Marel}}, \ and\ \bibinfo {author} {\bibfnamefont {C.~T.}\ \bibnamefont
  {Foxon}},\ }\href@noop {} {\bibfield  {journal} {\bibinfo  {journal} {Phys.
  Rev. Lett.}\ }\textbf {\bibinfo {volume} {60}},\ \bibinfo {pages} {848}
  (\bibinfo {year} {1988})}\BibitemShut {NoStop}%
\bibitem [{\citenamefont {Wharam}\ \emph {et~al.}(1988)\citenamefont {Wharam},
  \citenamefont {Thornton}, \citenamefont {Newbury}, \citenamefont {Pepper},
  \citenamefont {Ahmed}, \citenamefont {Frost}, \citenamefont {Hasko},
  \citenamefont {Peacock}, \citenamefont {Ritchie},\ and\ \citenamefont
  {Jones}}]{wharam:1988jphysc}%
  \BibitemOpen
  \bibfield  {author} {\bibinfo {author} {\bibfnamefont {D.~A.}\ \bibnamefont
  {Wharam}}, \bibinfo {author} {\bibfnamefont {T.~J.}\ \bibnamefont
  {Thornton}}, \bibinfo {author} {\bibfnamefont {R.}~\bibnamefont {Newbury}},
  \bibinfo {author} {\bibfnamefont {M.}~\bibnamefont {Pepper}}, \bibinfo
  {author} {\bibfnamefont {H.}~\bibnamefont {Ahmed}}, \bibinfo {author}
  {\bibfnamefont {J.~E.~F.}\ \bibnamefont {Frost}}, \bibinfo {author}
  {\bibfnamefont {D.~G.}\ \bibnamefont {Hasko}}, \bibinfo {author}
  {\bibfnamefont {D.~C.}\ \bibnamefont {Peacock}}, \bibinfo {author}
  {\bibfnamefont {D.~A.}\ \bibnamefont {Ritchie}}, \ and\ \bibinfo {author}
  {\bibfnamefont {G.~A.~C.}\ \bibnamefont {Jones}},\ }\href@noop {} {\bibfield
  {journal} {\bibinfo  {journal} {J. Phys. C}\ }\textbf {\bibinfo {volume}
  {21}},\ \bibinfo {pages} {L209} (\bibinfo {year} {1988})}\BibitemShut
  {NoStop}%
\bibitem [{\citenamefont {Flory}(1997)}]{flory:1997jap}%
  \BibitemOpen
  \bibfield  {author} {\bibinfo {author} {\bibfnamefont {C.~A.}\ \bibnamefont
  {Flory}},\ }\href@noop {} {\bibfield  {journal} {\bibinfo  {journal} {J.
  Appl. Phys.}\ }\textbf {\bibinfo {volume} {82}},\ \bibinfo {pages} {6306}
  (\bibinfo {year} {1997})}\BibitemShut {NoStop}%
\bibitem [{\citenamefont {Szafer}\ and\ \citenamefont
  {Stone}(1989)}]{szafer:1989prl}%
  \BibitemOpen
  \bibfield  {author} {\bibinfo {author} {\bibfnamefont {A.}~\bibnamefont
  {Szafer}}\ and\ \bibinfo {author} {\bibfnamefont {A.~D.}\ \bibnamefont
  {Stone}},\ }\href@noop {} {\bibfield  {journal} {\bibinfo  {journal} {Phys.
  Rev. Lett.}\ }\textbf {\bibinfo {volume} {62}},\ \bibinfo {pages} {300}
  (\bibinfo {year} {1989})}\BibitemShut {NoStop}%
\bibitem [{\citenamefont {van~der Marel}\ and\ \citenamefont
  {Haanappel}(1989)}]{marel:1989prb}%
  \BibitemOpen
  \bibfield  {author} {\bibinfo {author} {\bibfnamefont {D.}~\bibnamefont
  {van~der Marel}}\ and\ \bibinfo {author} {\bibfnamefont {E.~G.}\ \bibnamefont
  {Haanappel}},\ }\href@noop {} {\bibfield  {journal} {\bibinfo  {journal}
  {Phys. Rev. B}\ }\textbf {\bibinfo {volume} {39}},\ \bibinfo {pages} {7811}
  (\bibinfo {year} {1989})}\BibitemShut {NoStop}%
\bibitem [{\citenamefont {Tekman}\ and\ \citenamefont
  {Ciraci}(1991)}]{tekman:1991prb}%
  \BibitemOpen
  \bibfield  {author} {\bibinfo {author} {\bibfnamefont {E.}~\bibnamefont
  {Tekman}}\ and\ \bibinfo {author} {\bibfnamefont {S.}~\bibnamefont
  {Ciraci}},\ }\href@noop {} {\bibfield  {journal} {\bibinfo  {journal} {Phys.
  Rev. B}\ }\textbf {\bibinfo {volume} {43}},\ \bibinfo {pages} {7145}
  (\bibinfo {year} {1991})}\BibitemShut {NoStop}%
\bibitem [{\citenamefont {Yacoby}\ and\ \citenamefont
  {Imry}(1990)}]{yacoby:1990prb}%
  \BibitemOpen
  \bibfield  {author} {\bibinfo {author} {\bibfnamefont {A.}~\bibnamefont
  {Yacoby}}\ and\ \bibinfo {author} {\bibfnamefont {Y.}~\bibnamefont {Imry}},\
  }\href@noop {} {\bibfield  {journal} {\bibinfo  {journal} {Phys. Rev. B}\
  }\textbf {\bibinfo {volume} {41}},\ \bibinfo {pages} {5341} (\bibinfo {year}
  {1990})}\BibitemShut {NoStop}%
\bibitem [{\citenamefont {Sols}\ \emph {et~al.}(1989)\citenamefont {Sols},
  \citenamefont {Macucci}, \citenamefont {Ravaioli},\ and\ \citenamefont
  {Hess}}]{sols:1989apl}%
  \BibitemOpen
  \bibfield  {author} {\bibinfo {author} {\bibfnamefont {F.}~\bibnamefont
  {Sols}}, \bibinfo {author} {\bibfnamefont {M.}~\bibnamefont {Macucci}},
  \bibinfo {author} {\bibfnamefont {U.}~\bibnamefont {Ravaioli}}, \ and\
  \bibinfo {author} {\bibfnamefont {K.}~\bibnamefont {Hess}},\ }\href@noop {}
  {\bibfield  {journal} {\bibinfo  {journal} {Appl. Phys. Lett.}\ }\textbf
  {\bibinfo {volume} {54}},\ \bibinfo {pages} {350} (\bibinfo {year}
  {1989})}\BibitemShut {NoStop}%
\bibitem [{\citenamefont {Debray}\ \emph {et~al.}(2000)\citenamefont {Debray},
  \citenamefont {Raichev}, \citenamefont {Vasilopoulos}, \citenamefont
  {Rahman}, \citenamefont {Perrin},\ and\ \citenamefont
  {Mitchell}}]{debray:2000prb}%
  \BibitemOpen
  \bibfield  {author} {\bibinfo {author} {\bibfnamefont {P.}~\bibnamefont
  {Debray}}, \bibinfo {author} {\bibfnamefont {O.~E.}\ \bibnamefont {Raichev}},
  \bibinfo {author} {\bibfnamefont {P.}~\bibnamefont {Vasilopoulos}}, \bibinfo
  {author} {\bibfnamefont {M.}~\bibnamefont {Rahman}}, \bibinfo {author}
  {\bibfnamefont {R.}~\bibnamefont {Perrin}}, \ and\ \bibinfo {author}
  {\bibfnamefont {W.~C.}\ \bibnamefont {Mitchell}},\ }\href@noop {} {\bibfield
  {journal} {\bibinfo  {journal} {Phys. Rev. B}\ }\textbf {\bibinfo {volume}
  {61}},\ \bibinfo {pages} {10950} (\bibinfo {year} {2000})}\BibitemShut
  {NoStop}%
\bibitem [{\citenamefont {Kouwenhoven}\ \emph {et~al.}(1990)\citenamefont
  {Kouwenhoven}, \citenamefont {Hekking}, \citenamefont {van Wees},
  \citenamefont {Harmans}, \citenamefont {Timmering},\ and\ \citenamefont
  {Foxon}}]{kouwenhoven:1990prl}%
  \BibitemOpen
  \bibfield  {author} {\bibinfo {author} {\bibfnamefont {L.~P.}\ \bibnamefont
  {Kouwenhoven}}, \bibinfo {author} {\bibfnamefont {F.~W.~J.}\ \bibnamefont
  {Hekking}}, \bibinfo {author} {\bibfnamefont {B.~J.}\ \bibnamefont {van
  Wees}}, \bibinfo {author} {\bibfnamefont {C.~J. P.~M.}\ \bibnamefont
  {Harmans}}, \bibinfo {author} {\bibfnamefont {C.~E.}\ \bibnamefont
  {Timmering}}, \ and\ \bibinfo {author} {\bibfnamefont {C.~T.}\ \bibnamefont
  {Foxon}},\ }\href@noop {} {\bibfield  {journal} {\bibinfo  {journal} {Phys.
  Rev. Lett.}\ }\textbf {\bibinfo {volume} {65}},\ \bibinfo {pages} {361}
  (\bibinfo {year} {1990})}\BibitemShut {NoStop}%
\bibitem [{\citenamefont {Frost}\ \emph {et~al.}(1994)\citenamefont {Frost},
  \citenamefont {Liang}, \citenamefont {Pepper}, \citenamefont {Ritchie},
  \citenamefont {Linfield}, \citenamefont {Ford}, \citenamefont {Smith},\ and\
  \citenamefont {Jones}}]{frost:1994prb}%
  \BibitemOpen
  \bibfield  {author} {\bibinfo {author} {\bibfnamefont {I.~M.}\ \bibnamefont
  {Frost}, \bibfnamefont {J.~E. F~Castleton}}, \bibinfo {author} {\bibfnamefont
  {C.-T.}\ \bibnamefont {Liang}}, \bibinfo {author} {\bibfnamefont
  {M.}~\bibnamefont {Pepper}}, \bibinfo {author} {\bibfnamefont {D.~A.}\
  \bibnamefont {Ritchie}}, \bibinfo {author} {\bibfnamefont {E.~H.}\
  \bibnamefont {Linfield}}, \bibinfo {author} {\bibfnamefont {C.~J.~B.}\
  \bibnamefont {Ford}}, \bibinfo {author} {\bibfnamefont {C.~G.}\ \bibnamefont
  {Smith}}, \ and\ \bibinfo {author} {\bibfnamefont {G.~A.~C.}\ \bibnamefont
  {Jones}},\ }\href@noop {} {\bibfield  {journal} {\bibinfo  {journal} {Phys.
  Rev. B}\ }\textbf {\bibinfo {volume} {49}},\ \bibinfo {pages} {14078}
  (\bibinfo {year} {1994})}\BibitemShut {NoStop}%
\bibitem [{\citenamefont {Wen}\ \emph {et~al.}(2010)\citenamefont {Wen},
  \citenamefont {Hsiao}, \citenamefont {Lin}, \citenamefont {Hong},
  \citenamefont {Chen}, \citenamefont {Ueda},\ and\ \citenamefont
  {Komiyama}}]{wen:2010prb}%
  \BibitemOpen
  \bibfield  {author} {\bibinfo {author} {\bibfnamefont {C.-S.}\ \bibnamefont
  {Wen}}, \bibinfo {author} {\bibfnamefont {J.~H.}\ \bibnamefont {Hsiao}},
  \bibinfo {author} {\bibfnamefont {K.-T.}\ \bibnamefont {Lin}}, \bibinfo
  {author} {\bibfnamefont {T.-M.}\ \bibnamefont {Hong}}, \bibinfo {author}
  {\bibfnamefont {J.~C.}\ \bibnamefont {Chen}}, \bibinfo {author}
  {\bibfnamefont {T.}~\bibnamefont {Ueda}}, \ and\ \bibinfo {author}
  {\bibfnamefont {S.}~\bibnamefont {Komiyama}},\ }\href@noop {} {\bibfield
  {journal} {\bibinfo  {journal} {Phys. Rev. B}\ }\textbf {\bibinfo {volume}
  {82}},\ \bibinfo {pages} {115416} (\bibinfo {year} {2010})}\BibitemShut
  {NoStop}%
\bibitem [{\citenamefont {Wang}\ \emph {et~al.}(2000)\citenamefont {Wang},
  \citenamefont {Carlsson}, \citenamefont {Maximov}, \citenamefont {Omling},
  \citenamefont {Samuelson}, \citenamefont {Seifert}, \citenamefont {Sheng},
  \citenamefont {Shorubalko},\ and\ \citenamefont {Xu}}]{wang:2000apl}%
  \BibitemOpen
  \bibfield  {author} {\bibinfo {author} {\bibfnamefont {Q.}~\bibnamefont
  {Wang}}, \bibinfo {author} {\bibfnamefont {N.}~\bibnamefont {Carlsson}},
  \bibinfo {author} {\bibfnamefont {I.}~\bibnamefont {Maximov}}, \bibinfo
  {author} {\bibfnamefont {P.}~\bibnamefont {Omling}}, \bibinfo {author}
  {\bibfnamefont {L.}~\bibnamefont {Samuelson}}, \bibinfo {author}
  {\bibfnamefont {W.}~\bibnamefont {Seifert}}, \bibinfo {author} {\bibfnamefont
  {W.}~\bibnamefont {Sheng}}, \bibinfo {author} {\bibfnamefont
  {I.}~\bibnamefont {Shorubalko}}, \ and\ \bibinfo {author} {\bibfnamefont
  {H.~Q.}\ \bibnamefont {Xu}},\ }\href@noop {} {\bibfield  {journal} {\bibinfo
  {journal} {Appl. Phys. Lett.}\ }\textbf {\bibinfo {volume} {76}},\ \bibinfo
  {pages} {2274} (\bibinfo {year} {2000})}\BibitemShut {NoStop}%
\bibitem [{\citenamefont {Liang}\ \emph {et~al.}(2001)\citenamefont {Liang},
  \citenamefont {Bockrath}, \citenamefont {Bozovic}, \citenamefont {Hafner},
  \citenamefont {Tinkham},\ and\ \citenamefont {Park}}]{liang:2001nature}%
  \BibitemOpen
  \bibfield  {author} {\bibinfo {author} {\bibfnamefont {W.}~\bibnamefont
  {Liang}}, \bibinfo {author} {\bibfnamefont {M.}~\bibnamefont {Bockrath}},
  \bibinfo {author} {\bibfnamefont {D.}~\bibnamefont {Bozovic}}, \bibinfo
  {author} {\bibfnamefont {J.~H.}\ \bibnamefont {Hafner}}, \bibinfo {author}
  {\bibfnamefont {M.}~\bibnamefont {Tinkham}}, \ and\ \bibinfo {author}
  {\bibfnamefont {H.}~\bibnamefont {Park}},\ }\href@noop {} {\bibfield
  {journal} {\bibinfo  {journal} {Nature}\ }\textbf {\bibinfo {volume} {411}},\
  \bibinfo {pages} {665} (\bibinfo {year} {2001})}\BibitemShut {NoStop}%
\bibitem [{\citenamefont {Kretinin}\ \emph {et~al.}(2010)\citenamefont
  {Kretinin}, \citenamefont {P.-Biro}, \citenamefont {Mahalu},\ and\
  \citenamefont {Shtrikman}}]{kretinin:2010nanolett}%
  \BibitemOpen
  \bibfield  {author} {\bibinfo {author} {\bibfnamefont {A.~V.}\ \bibnamefont
  {Kretinin}}, \bibinfo {author} {\bibfnamefont {R.}~\bibnamefont {P.-Biro}},
  \bibinfo {author} {\bibfnamefont {D.}~\bibnamefont {Mahalu}}, \ and\ \bibinfo
  {author} {\bibfnamefont {H.}~\bibnamefont {Shtrikman}},\ }\href@noop {}
  {\bibfield  {journal} {\bibinfo  {journal} {Nano lett.}\ }\textbf {\bibinfo
  {volume} {10}},\ \bibinfo {pages} {3439} (\bibinfo {year}
  {2010})}\BibitemShut {NoStop}%
\bibitem [{\citenamefont {Chou}\ and\ \citenamefont
  {Wang}(1993)}]{chou:1993apl}%
  \BibitemOpen
  \bibfield  {author} {\bibinfo {author} {\bibfnamefont {S.~Y.}\ \bibnamefont
  {Chou}}\ and\ \bibinfo {author} {\bibfnamefont {Y.}~\bibnamefont {Wang}},\
  }\href@noop {} {\bibfield  {journal} {\bibinfo  {journal} {Appl. phys.
  lett.}\ }\textbf {\bibinfo {volume} {63}},\ \bibinfo {pages} {788} (\bibinfo
  {year} {1993})}\BibitemShut {NoStop}%
\bibitem [{\citenamefont {Lee}\ \emph {et~al.}(2006)\citenamefont {Lee},
  \citenamefont {Muraki}, \citenamefont {Chang},\ and\ \citenamefont
  {Hirayama}}]{lee:2006jap}%
  \BibitemOpen
  \bibfield  {author} {\bibinfo {author} {\bibfnamefont {H.-M.}\ \bibnamefont
  {Lee}}, \bibinfo {author} {\bibfnamefont {K.}~\bibnamefont {Muraki}},
  \bibinfo {author} {\bibfnamefont {E.~Y.}\ \bibnamefont {Chang}}, \ and\
  \bibinfo {author} {\bibfnamefont {Y.}~\bibnamefont {Hirayama}},\ }\href@noop
  {} {\bibfield  {journal} {\bibinfo  {journal} {J. Appl. Phys.}\ }\textbf
  {\bibinfo {volume} {100}} (\bibinfo {year} {2006})}\BibitemShut {NoStop}%
\bibitem [{\citenamefont {Harrang}\ \emph {et~al.}(1985)\citenamefont
  {Harrang}, \citenamefont {Higgins}, \citenamefont {Goodall}, \citenamefont
  {Jay}, \citenamefont {Laviron},\ and\ \citenamefont
  {Delescluse}}]{harrang:1985prb}%
  \BibitemOpen
  \bibfield  {author} {\bibinfo {author} {\bibfnamefont {J.~P.}\ \bibnamefont
  {Harrang}}, \bibinfo {author} {\bibfnamefont {R.~J.}\ \bibnamefont
  {Higgins}}, \bibinfo {author} {\bibfnamefont {R.~K.}\ \bibnamefont
  {Goodall}}, \bibinfo {author} {\bibfnamefont {P.~R.}\ \bibnamefont {Jay}},
  \bibinfo {author} {\bibfnamefont {M.}~\bibnamefont {Laviron}}, \ and\
  \bibinfo {author} {\bibfnamefont {P.}~\bibnamefont {Delescluse}},\
  }\href@noop {} {\bibfield  {journal} {\bibinfo  {journal} {Phys. Rev. B}\
  }\textbf {\bibinfo {volume} {32}},\ \bibinfo {pages} {8126} (\bibinfo {year}
  {1985})}\BibitemShut {NoStop}%
\bibitem [{\citenamefont {Hamilton}\ \emph {et~al.}(1992)\citenamefont
  {Hamilton}, \citenamefont {Frost}, \citenamefont {Smith}, \citenamefont
  {Kelly}, \citenamefont {Linfield}, \citenamefont {Ford}, \citenamefont
  {Ritchie}, \citenamefont {Jones}, \citenamefont {Pepper}, \citenamefont
  {Hasko},\ and\ \citenamefont {Ahmed}}]{hamilton:1992apl}%
  \BibitemOpen
  \bibfield  {author} {\bibinfo {author} {\bibfnamefont {A.~R.}\ \bibnamefont
  {Hamilton}}, \bibinfo {author} {\bibfnamefont {J.~E.~F.}\ \bibnamefont
  {Frost}}, \bibinfo {author} {\bibfnamefont {C.~G.}\ \bibnamefont {Smith}},
  \bibinfo {author} {\bibfnamefont {M.~J.}\ \bibnamefont {Kelly}}, \bibinfo
  {author} {\bibfnamefont {E.~H.}\ \bibnamefont {Linfield}}, \bibinfo {author}
  {\bibfnamefont {C.~J.~B.}\ \bibnamefont {Ford}}, \bibinfo {author}
  {\bibfnamefont {D.~A.}\ \bibnamefont {Ritchie}}, \bibinfo {author}
  {\bibfnamefont {G.~A.~C.}\ \bibnamefont {Jones}}, \bibinfo {author}
  {\bibfnamefont {M.}~\bibnamefont {Pepper}}, \bibinfo {author} {\bibfnamefont
  {D.~G.}\ \bibnamefont {Hasko}}, \ and\ \bibinfo {author} {\bibfnamefont
  {H.}~\bibnamefont {Ahmed}},\ }\href@noop {} {\bibfield  {journal} {\bibinfo
  {journal} {Appl. Phys. Lett.}\ }\textbf {\bibinfo {volume} {60}},\ \bibinfo
  {pages} {2782} (\bibinfo {year} {1992})}\BibitemShut {NoStop}%
\bibitem [{\citenamefont {MacLeod}\ \emph {et~al.}(2009)\citenamefont
  {MacLeod}, \citenamefont {Chan}, \citenamefont {Martin}, \citenamefont
  {Hamilton}, \citenamefont {See}, \citenamefont {Micolich}, \citenamefont
  {Aagesen},\ and\ \citenamefont {Lindelof}}]{macleod:2009prb}%
  \BibitemOpen
  \bibfield  {author} {\bibinfo {author} {\bibfnamefont {S.~J.}\ \bibnamefont
  {MacLeod}}, \bibinfo {author} {\bibfnamefont {K.}~\bibnamefont {Chan}},
  \bibinfo {author} {\bibfnamefont {T.~P.}\ \bibnamefont {Martin}}, \bibinfo
  {author} {\bibfnamefont {A.~R.}\ \bibnamefont {Hamilton}}, \bibinfo {author}
  {\bibfnamefont {A.}~\bibnamefont {See}}, \bibinfo {author} {\bibfnamefont
  {A.~P.}\ \bibnamefont {Micolich}}, \bibinfo {author} {\bibfnamefont
  {M.}~\bibnamefont {Aagesen}}, \ and\ \bibinfo {author} {\bibfnamefont
  {P.~E.}\ \bibnamefont {Lindelof}},\ }\href@noop {} {\bibfield  {journal}
  {\bibinfo  {journal} {Phys. Rev. B}\ }\textbf {\bibinfo {volume} {80}},\
  \bibinfo {pages} {035310} (\bibinfo {year} {2009})}\BibitemShut {NoStop}%
\bibitem [{\citenamefont {Thomas}\ \emph {et~al.}(1996)\citenamefont {Thomas},
  \citenamefont {Nicholls}, \citenamefont {Simmons}, \citenamefont {Pepper},
  \citenamefont {Mace},\ and\ \citenamefont {Ritchie}}]{thomas:1996prl}%
  \BibitemOpen
  \bibfield  {author} {\bibinfo {author} {\bibfnamefont {K.~J.}\ \bibnamefont
  {Thomas}}, \bibinfo {author} {\bibfnamefont {J.~T.}\ \bibnamefont
  {Nicholls}}, \bibinfo {author} {\bibfnamefont {M.~Y.}\ \bibnamefont
  {Simmons}}, \bibinfo {author} {\bibfnamefont {M.}~\bibnamefont {Pepper}},
  \bibinfo {author} {\bibfnamefont {D.~R.}\ \bibnamefont {Mace}}, \ and\
  \bibinfo {author} {\bibfnamefont {D.~A.}\ \bibnamefont {Ritchie}},\
  }\href@noop {} {\bibfield  {journal} {\bibinfo  {journal} {Phys. Rev. Lett.}\
  }\textbf {\bibinfo {volume} {77}},\ \bibinfo {pages} {135} (\bibinfo {year}
  {1996})}\BibitemShut {NoStop}%
\bibitem [{\citenamefont {Heyder}\ \emph {et~al.}(2015)\citenamefont {Heyder},
  \citenamefont {Bauer}, \citenamefont {Schubert}, \citenamefont {Borowsky},
  \citenamefont {Schuh}, \citenamefont {Wegscheider}, \citenamefont {von
  Delft},\ and\ \citenamefont {Ludwig}}]{heyder:2015prb}%
  \BibitemOpen
  \bibfield  {author} {\bibinfo {author} {\bibfnamefont {J.}~\bibnamefont
  {Heyder}}, \bibinfo {author} {\bibfnamefont {F.}~\bibnamefont {Bauer}},
  \bibinfo {author} {\bibfnamefont {E.}~\bibnamefont {Schubert}}, \bibinfo
  {author} {\bibfnamefont {D.}~\bibnamefont {Borowsky}}, \bibinfo {author}
  {\bibfnamefont {D.}~\bibnamefont {Schuh}}, \bibinfo {author} {\bibfnamefont
  {W.}~\bibnamefont {Wegscheider}}, \bibinfo {author} {\bibfnamefont
  {J.}~\bibnamefont {von Delft}}, \ and\ \bibinfo {author} {\bibfnamefont
  {S.}~\bibnamefont {Ludwig}},\ }\href@noop {} {\bibfield  {journal} {\bibinfo
  {journal} {Phys. Rev. B}\ }\textbf {\bibinfo {volume} {92}},\ \bibinfo
  {pages} {195401} (\bibinfo {year} {2015})}\BibitemShut {NoStop}%
\bibitem [{\citenamefont {Reilly}\ \emph {et~al.}(2001)\citenamefont {Reilly},
  \citenamefont {Facer}, \citenamefont {Dzurak}, \citenamefont {Kane},
  \citenamefont {Clark}, \citenamefont {Stiles}, \citenamefont {Hamilton},
  \citenamefont {O'Brien}, \citenamefont {Lumpkin}, \citenamefont {Pfeiffer},\
  and\ \citenamefont {West}}]{reilly:2001prb}%
  \BibitemOpen
  \bibfield  {author} {\bibinfo {author} {\bibfnamefont {D.~J.}\ \bibnamefont
  {Reilly}}, \bibinfo {author} {\bibfnamefont {G.~R.}\ \bibnamefont {Facer}},
  \bibinfo {author} {\bibfnamefont {A.~S.}\ \bibnamefont {Dzurak}}, \bibinfo
  {author} {\bibfnamefont {B.~E.}\ \bibnamefont {Kane}}, \bibinfo {author}
  {\bibfnamefont {R.~G.}\ \bibnamefont {Clark}}, \bibinfo {author}
  {\bibfnamefont {P.~J.}\ \bibnamefont {Stiles}}, \bibinfo {author}
  {\bibfnamefont {A.~R.}\ \bibnamefont {Hamilton}}, \bibinfo {author}
  {\bibfnamefont {J.~L.}\ \bibnamefont {O'Brien}}, \bibinfo {author}
  {\bibfnamefont {N.~E.}\ \bibnamefont {Lumpkin}}, \bibinfo {author}
  {\bibfnamefont {L.~N.}\ \bibnamefont {Pfeiffer}}, \ and\ \bibinfo {author}
  {\bibfnamefont {K.~W.}\ \bibnamefont {West}},\ }\href@noop {} {\bibfield
  {journal} {\bibinfo  {journal} {Phys. Rev. B}\ }\textbf {\bibinfo {volume}
  {63}},\ \bibinfo {pages} {121311} (\bibinfo {year} {2001})}\BibitemShut
  {NoStop}%
\bibitem [{\citenamefont {Smith}\ \emph {et~al.}(2016)\citenamefont {Smith},
  \citenamefont {Al-Taie}, \citenamefont {Lesage}, \citenamefont {Thomas},
  \citenamefont {Sfigakis}, \citenamefont {See}, \citenamefont {Griffiths},
  \citenamefont {Farrer}, \citenamefont {Jones}, \citenamefont {Ritchie},
  \citenamefont {Kelly},\ and\ \citenamefont {Smith}}]{smith:2016prappl}%
  \BibitemOpen
  \bibfield  {author} {\bibinfo {author} {\bibfnamefont {L.~W.}\ \bibnamefont
  {Smith}}, \bibinfo {author} {\bibfnamefont {H.}~\bibnamefont {Al-Taie}},
  \bibinfo {author} {\bibfnamefont {A.~A.~J.}\ \bibnamefont {Lesage}}, \bibinfo
  {author} {\bibfnamefont {K.~J.}\ \bibnamefont {Thomas}}, \bibinfo {author}
  {\bibfnamefont {F.}~\bibnamefont {Sfigakis}}, \bibinfo {author}
  {\bibfnamefont {P.}~\bibnamefont {See}}, \bibinfo {author} {\bibfnamefont
  {J.~P.}\ \bibnamefont {Griffiths}}, \bibinfo {author} {\bibfnamefont
  {I.}~\bibnamefont {Farrer}}, \bibinfo {author} {\bibfnamefont {G.~A.~C.}\
  \bibnamefont {Jones}}, \bibinfo {author} {\bibfnamefont {D.~A.}\ \bibnamefont
  {Ritchie}}, \bibinfo {author} {\bibfnamefont {M.~J.}\ \bibnamefont {Kelly}},
  \ and\ \bibinfo {author} {\bibfnamefont {C.~G.}\ \bibnamefont {Smith}},\
  }\href@noop {} {\bibfield  {journal} {\bibinfo  {journal} {Phys. Rev. Appl.}\
  }\textbf {\bibinfo {volume} {5}},\ \bibinfo {pages} {044015} (\bibinfo {year}
  {2016})}\BibitemShut {NoStop}%
\bibitem [{\citenamefont {Kristensen}\ \emph {et~al.}(2000)\citenamefont
  {Kristensen}, \citenamefont {Bruus}, \citenamefont {Hansen}, \citenamefont
  {Jensen}, \citenamefont {Lindelof}, \citenamefont {Marckmann}, \citenamefont
  {Nyg{\aa}rd}, \citenamefont {S{\o}rensen}, \citenamefont {Beuscher},
  \citenamefont {Forchel},\ and\ \citenamefont {Michel}}]{kristensen:2000prb}%
  \BibitemOpen
  \bibfield  {author} {\bibinfo {author} {\bibfnamefont {A.}~\bibnamefont
  {Kristensen}}, \bibinfo {author} {\bibfnamefont {H.}~\bibnamefont {Bruus}},
  \bibinfo {author} {\bibfnamefont {A.~E.}\ \bibnamefont {Hansen}}, \bibinfo
  {author} {\bibfnamefont {J.~B.}\ \bibnamefont {Jensen}}, \bibinfo {author}
  {\bibfnamefont {P.~E.}\ \bibnamefont {Lindelof}}, \bibinfo {author}
  {\bibfnamefont {C.~J.}\ \bibnamefont {Marckmann}}, \bibinfo {author}
  {\bibfnamefont {J.}~\bibnamefont {Nyg{\aa}rd}}, \bibinfo {author}
  {\bibfnamefont {C.~B.}\ \bibnamefont {S{\o}rensen}}, \bibinfo {author}
  {\bibfnamefont {F.}~\bibnamefont {Beuscher}}, \bibinfo {author}
  {\bibfnamefont {A.}~\bibnamefont {Forchel}}, \ and\ \bibinfo {author}
  {\bibfnamefont {M.}~\bibnamefont {Michel}},\ }\href@noop {} {\bibfield
  {journal} {\bibinfo  {journal} {Phys. Rev. B}\ }\textbf {\bibinfo {volume}
  {62}},\ \bibinfo {pages} {10950} (\bibinfo {year} {2000})}\BibitemShut
  {NoStop}%
\bibitem [{\citenamefont {Iqbal}\ \emph {et~al.}(2013)\citenamefont {Iqbal},
  \citenamefont {de. Jong}, \citenamefont {Reuter}, \citenamefont {Wieck},\
  and\ \citenamefont {van~der Wal}}]{iqbal:2013jap}%
  \BibitemOpen
  \bibfield  {author} {\bibinfo {author} {\bibfnamefont {M.~J.}\ \bibnamefont
  {Iqbal}}, \bibinfo {author} {\bibfnamefont {J.~P.}\ \bibnamefont {de. Jong}},
  \bibinfo {author} {\bibfnamefont {D.}~\bibnamefont {Reuter}}, \bibinfo
  {author} {\bibfnamefont {A.~D.}\ \bibnamefont {Wieck}}, \ and\ \bibinfo
  {author} {\bibfnamefont {C.~H.}\ \bibnamefont {van~der Wal}},\ }\href@noop {}
  {\bibfield  {journal} {\bibinfo  {journal} {J. Appl. Phys.}\ }\textbf
  {\bibinfo {volume} {113}},\ \bibinfo {pages} {024507} (\bibinfo {year}
  {2013})}\BibitemShut {NoStop}%
\end{thebibliography}
%

\end{document}